\documentclass[aps,pra,final,10pt,twocolumn,superscriptaddress]{revtex4-1}
\usepackage{graphicx}
\usepackage{amsmath}
\usepackage{dcolumn}
\usepackage{bm}
\usepackage{ulem}
\usepackage{times}
\usepackage{amssymb}
\usepackage{epsfig}
\usepackage{xcolor}

\begin{document}
\title{Aging and rejuvenation of active matter under topological constraints}


\author{Liesbeth M.~C.~Janssen}
\email[Electronic mail: ]{ljanssen@thphy.uni-duesseldorf.de}
\affiliation{Institute for Theoretical Physics II: Soft Matter, Heinrich-Heine University D\"{u}sseldorf,
             Universit\"{a}tsstra{\ss}e 1, 40225 D\"{u}sseldorf, Germany}
\author{Andreas Kaiser}
\affiliation{Materials Science Division, Argonne National Laboratory, 9700 South Cass Av, Illinois 60439, USA}
\author{Hartmut L\"{o}wen}
\affiliation{Institute for Theoretical Physics II: Soft Matter, Heinrich-Heine University D\"{u}sseldorf,
             Universit\"{a}tsstra{\ss}e 1, 40225 D\"{u}sseldorf, Germany}

\date{\today}

\maketitle


\textbf{
The coupling of active, self-motile particles to topological constraints can
give rise to novel non-equilibrium dynamical patterns that lack any passive
counterpart. Here we study the behavior of self-propelled rods confined to a compact
spherical manifold by means of Brownian dynamics simulations. We establish the
state diagram and find that short active rods at sufficiently high density exhibit a
glass transition toward a disordered state characterized by persistent
self-spinning motion. By periodically melting and revitrifying the spherical
spinning glass, we observe clear signatures of time-dependent aging and
rejuvenation physics. We quantify the crucial role of activity in these
non-equilibrium processes, and rationalize the aging dynamics in terms of an
absorbing-state transition toward a more stable active glassy state. Our results
demonstrate both how concepts of passive glass phenomenology can carry over into 
the realm of active matter, and how topology can enrich the collective
spatiotemporal dynamics in inherently non-equilibrium systems.
}

Active systems are composed of particles that can convert chemical, magnetic,
or radiation energy into autonomous motion, rendering them intrinsically far
from equilibrium \cite{Marchetti2013a,Romanczuk2012,Elgeti2015,Zottl2016}. Examples of living
active matter are found on all length scales, from microscopic motile bacteria
to macroscopic flocks of birds, and also numerous synthetic active materials
have recently become available \cite{Bechinger2016}.  The spatiotemporal
dynamics exhibited by such systems range from swarming and giant number
fluctuations \cite{Vicsek1995,Narayan2007} to low-Reynolds-number turbulence
\cite{Wensink2012,Rabani2013,Bratanov2015,Giomi2015} and motility-induced phase
separation \cite{Tailleur2008,Fily2012,Redner2013,Buttinoni2013}, illustrating
the rich collective behavior that emerges from the non-equilibrium energy
dissipation and active self-motility at the single-particle level.

It was recently found that sufficiently dense assemblies of active matter can
also exhibit hallmarks of glassy dynamics
\cite{Henkes2011,Angelini2011,Ni2013,Berthier2014, Pilkiewicz2014, Szamel2015,
Szamel2015a, Farage2014, Ding2015, Bi2015, Bi2016, Delarue2016, Yazdi2016}, including slow relaxation,
dynamic heterogeneity, and ultimate kinetic arrest--akin to the behavior
observed in non-active supercooled liquids and dense colloidal suspensions
\cite{Berthier2011a}. For passive systems, the process of glass formation has
been widely studied over the last few decades, resulting in multiple compelling
theoretical scenarios for the conventional glass transition
\cite{Berthier2011a,Biroli2013,Royall2015}. However, the extent to which the
phenomenology of passive glass-formers differs from that of non-equilibium
dense active matter remains a topic of scientific debate. For example, while
initial simulation studies suggested that adding activity generally pushes the
glass transition to higher densities and lower temperatures
\cite{Ni2013,Berthier2014}, more recent work argues that active self-motility
can both increase and decrease a system's glassiness \cite{Szamel2015a,
Szamel2015}.  This indicates that activity has a more intricate effect on the
dynamics than merely shifting the effective density or temperature. The
question whether time-dependent out-of-equilibrium glassy phenomena such as aging and
rejuvenation may also occur in active matter has thus far remained unexplored.  
These latter processes are generally understood in terms of an energy-landscape picture,
whereby aging and rejuvenation correspond to relaxation toward deeper and shallower
energy minima, respectively \cite{Berthier2009}. However, owing to the non-Hamiltonian nature of particle activity,
the potential (or free) energy is generally not a useful metric for active matter, and hence  
it remains unclear if and how aging and rejuvenation might be manifested in an active glass.

A different avenue of research concerns the effects of geometric 
\cite{Bricard2013, Koumakis2013, Das2015, Simmchen2016, Morin2016,Takatori2016}
and topological \cite{Keber2014, Grossmann2015,Li2015,Fily2016,Sknepnek2015,Khoromskaia2016,Ehrig2016}
constraints on active matter. For passive soft matter systems, it
is well established that confining a system to a curved surface can both
frustrate and promote long-range orientational order
\cite{Berreman1972,Lopez-Leon2011, Brito2016}, induce complex
topological-defect structures
\cite{Fernandez-Nieves2007,Backofen2010,Irvine2010,Irvine2012}, and affect a system's
glass-forming properties \cite{Sausset2008}.  In a biological context, surface
curvature is known to play a role in collective cell migration during e.g.\
embryonic development \cite{Keller2008} and the growth of the corneal
epithelium \cite{Collinson2002}.  For active soft-matter systems, however, only
a limited number of experimental and theoretical studies has addressed the role
of curvature and topology. Explicitly, recent experimental work has focused on
active nematic microtubuli confined to a deformable droplet interface \cite{Keber2014},
and subsequent theoretical \cite{Grossmann2015,Li2015,Fily2016} and simulation
\cite{Sknepnek2015,Khoromskaia2016,Ehrig2016} studies have explored the
dynamics of nematic and polar active particles under a spherical or ellipsoidal
constraint.  These developments point toward a rich array of topological-defect
patterns and curvature-driven dynamics in the presence of strong aligning
interactions between the particles.  It remains unclear, however, how active
particles with weak alignment interactions behave under topological
constraints, and how disordered glass-like dynamics may possibly emerge under
such conditions.

Here we seek to unite these independent lines of research and present a
systematic study of the interplay between topology, particle activity, and
effective particle alignment interactions. Specifically, we perform Brownian
dynamics simulations of repulsive, self-propelled polar rods confined to a
compact spherical manifold, and explore the emergent collective dynamics for
different packing densities and particle aspect ratios. We find that
particularly the high-density regimes are
influenced by the confining topology, and for sufficiently dense short rods, we
observe a novel glass transition toward a solid-like disordered state in which
all particles undergo collective rotation. Remarkably,
upon repeated melting and vitrification of this self-spinning glass phase, we
also find evidence of aging and rejuvenation dynamics, which we clarify in terms
of an absorbing-state formalism and a stability-landscape
picture. Overall, our results exemplify both the novel spatiotemporal dynamics
that may emerge from coupling activity to topology, and the surprising
analogies between active matter that is intrinsically out-of-equilibrium, and
passive glassy matter that is collectively out-of-equilibrium.  We expect our
findings to be verifiable in experiments on e.g.\ dense suspensions of
bacterial or synthetic active particles confined to a spherical droplet or
hydrogel interface.

\section{Results} 

\subsection{State diagram} 
\begin{figure*}
        \begin{center}
    \includegraphics[width=0.9\textwidth]{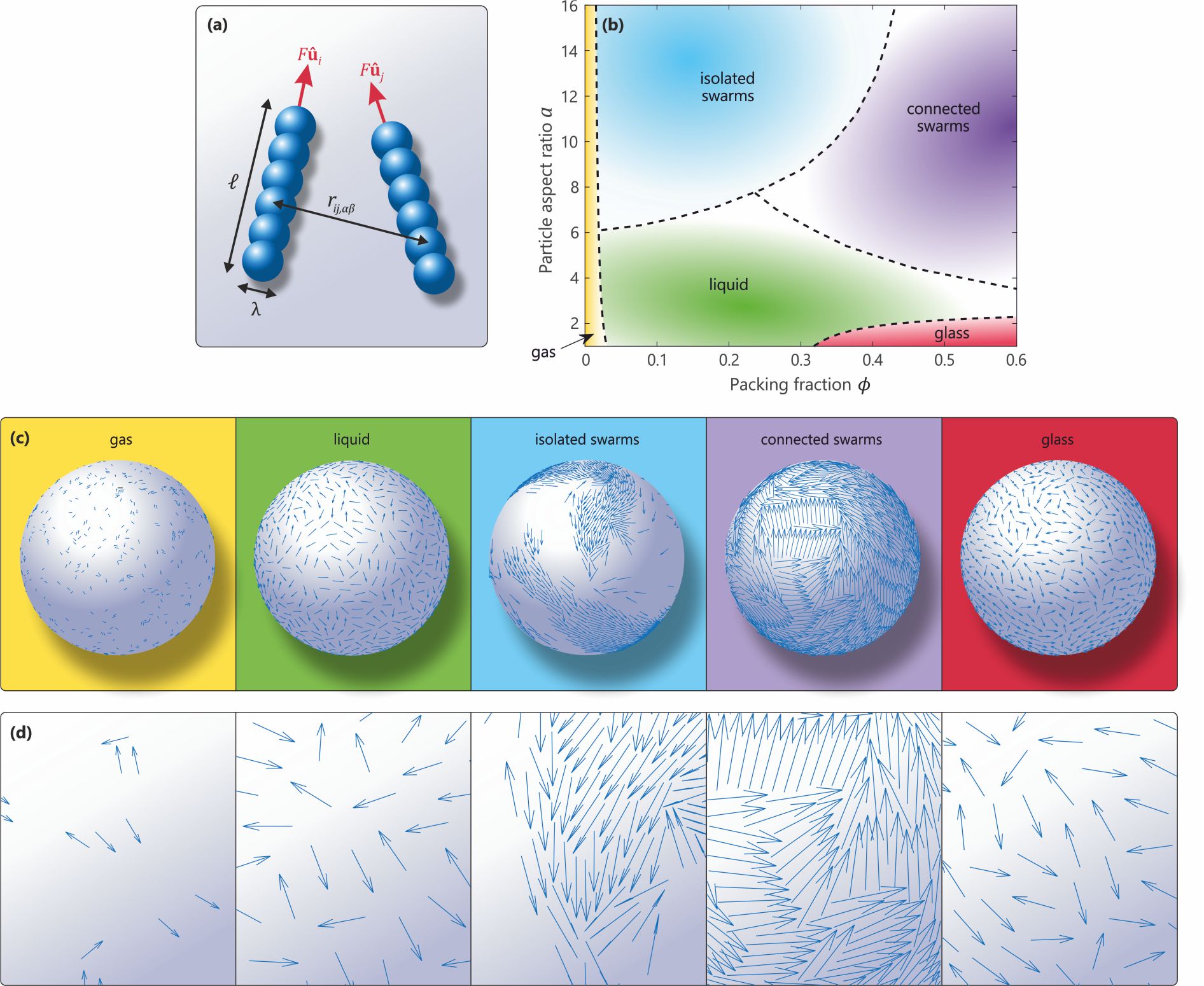}
  \end{center}
  \caption{
  \label{fig:statediagram} 
\textbf{Non-equilibrium state diagram for active rods on a sphere.}
(a) Schematic representation of our active-rod model system. 
(b) State diagram for $N=800$ active particles on a spherical surface
as a function of the particle aspect ratio $a=\ell/\lambda$ and packing fraction $\phi$.
The different phases were identified by visual inspection of each indivual trajectory. 
Dashed lines indicate approximate boundaries between phases and serve as a guide to the eye. 
The evaluated state points are indicated in the Supplementary Information.
(c) Typical snapshots of the different phases indicated in the state diagram:
gas ($a=10, \phi=0.01$), liquid ($a=4, \phi=0.2$), isolated swarms ($a=16,
\phi=0.1$), connected swarms ($a=10, \phi=0.6$), and glass ($a=2, \phi=0.5$).
(d) Corresponding close-ups of the snapshots.  Every blue arrow represents a
single particle with orientation vector $\hat{\mathbf{u}}_i$.
  }
\end{figure*}

We first explore the full non-equilibrium state diagram of self-propelled rods
on a sphere as a function of the packing fraction and particle aspect ratio.
Our system is based on a suitable minimal model system for bacterial
microswimmers in Euclidean space \cite{Wensink2012,Wensink2012a}, which is
illustrated in Fig.\ \ref{fig:statediagram}(a) and discussed in detail in the
Methods section. Briefly, we consider $N$ rigid, self-propelled rods of length
$\ell$ that move with a constant self-propulsion force $F$ directed along the
main rod axis $\hat{\mathbf{u}}$. Each rod consists of $n$ spherical segments
that interact with the segments of any other rod through a steeply repulsive
Yukawa potential, preventing particles to overlap. The screening length
$\lambda$ of the Yukawa interaction defines the effective width of the rods and
serves as our unit of length. We perform a series of overdamped Brownian
dynamics simulations as a function of the particle aspect ratio
$a=\ell/\lambda$ and effective packing fraction $\phi = N\ell\lambda/(4\pi
R^2)$, where $R$ denotes the radius of the confining sphere.  Throughout the
simulations, the rods are constrained to lie tangent to the surface of the
confining spherical manifold, with each rod's center-of-mass position
$\mathbf{r}_i$ connected to the sphere. For simplicity we ignore hydrodynamic
interactions and thermal noise, thus allowing us to focus on a minimal model
system that captures the interplay between the particles' geometry, packing
density, and topology of the confining sphere.  Finally, considering the
inherent finize size of a spherical surface, which implicitly prevents the
existence of a thermodynamic limit, we restrict ourselves to the behavior of
small systems of typically 400-800 particles.

Figure \ref{fig:statediagram}(b) shows the state diagram of our spherically
constrained active-rod system as a function of the rod aspect ratio $a$ and
packing fraction $\phi$, calculated for a system of $N=800$ particles.  
Snapshots of the corresponding phases are shown in Figs.\
\ref{fig:statediagram}(c) and (d), and the time-dependent dynamics can be
seen in Supplementary Movies S1 to S5. With the exception of extremely dilute
packings $\phi \lesssim 0.01$--in which case an active-gas phase forms--we can
identify a marked dependence on rod length in the dynamical behavior.  For
\textit{large} particle aspect ratios, we find that the rods tend to align and
spontaneously form domains of local polar order. This alignment effect is well
established for active repulsive rods in 2D Euclidean space, and here we find
that it also applies in curved space. The observed alignment is the result of
pair collisions: when two active rods collide, the resulting torques and steric
forces cause the rods to orient in the same direction and move close to each
other--even though no attractive forces exist between the particles
\cite{Weitz2015}.  At low packing fractions, this leads to a distinct
\textit{swarming} phase in which the rods group together in isolated flocks and
exhibit giant density fluctuations, completely analogous to swarming in
Euclidean space \cite{Wensink2012a}. For higher packing fractions, however, the
rods experience a packing constraint and become affected by the presence of the
confining topology: the different swarms become connected and form a giant,
dynamic ''multi-domain swarm" that ultimately spans the entire sphere.  As in
the lower-density swarming phase, each of these domains is composed of locally
oriented rods with polar and/or smectic order.  At sufficiently high densities,
transient topological defects can be identified at the boundaries between the
different domains, and the dynamics becomes a rich pattern of mobile defects
and transient counter flows. 

We remark that in the limit of an infinite sphere radius (or infinite particle number $N$),
our state diagram should extrapolate to that for a flat 2D surface. The latter
contains a distinct turbulent and laning phase for long rods at high density \cite{Wensink2012a},
while in our current work we can identify only a "connected swarms" phase. 
The fact that we do not observe a well-developed turbulent phase here is likely
due to the relatively small number of particles used, preventing the formation
of a coarse-grained vorticity field. However, the fact that we do not observe a
clear laning phase is inherently due to the confining topology: at least for small
system sizes of $N=800$ rods, we have verified that a flat 2D surface with periodic boundary conditions
quickly gives rise to distinct laning, while on the sphere such a phase is never
stable. Thus, if the system size is sufficiently small to "feel" the presence
of the confining spherical topology, the 2D global laning phase is destabilized
and converted into the dynamic "connected swarms" phase that exhibits only
local and transient laning-like behavior.

The emergent dynamics becomes dramatically different when reducing the particle
aspect ratio $a$. Short rods experience only a small torque during a pair
collision, causing the alignment effect to eventually vanish and consequently
giving rise to strongly disordered dynamics.  Indeed, for short rods at low
packing fractions, we observe an active-liquid phase in which the particles
move incoherently and exhibit no strong cooperative motion. Note that in this
state, in contrast to the long-rod case, the particles are all oriented
randomly and are spread homogeneously across the surface of the sphere.

Intriguingly, we find that at sufficiently dense packings, systems with $a<2.5$
undergo a marked transition into a kinetically arrested state, as depicted in
Fig.\ \ref{fig:statediagram} and Supplementary Movie S5. In this non-ergodic phase, which we
term a \textit{self-spinning glass}, the relative positions and orientations of
all particles are frozen in a disordered configuration, but the system as a
whole undergoes a \textit{collective} rotation about a fixed arbitrary axis
with constant angular velocity.  The source of this self-sustained spinning
dynamics lies in the activity: every particle in the glassy state exerts a
constant self-propulsion force $F$ in a (quasi-)random direction, giving rise
to a net (random) force that in general will be small but nonzero.  This, in
turn, produces a finite torque that drives the collective rotation. Note that
such a spinning motion is a consequence of the unique topology of the sphere
and would be unattainable on a flat 2D plane--the latter permitting only
collective translational motion, as indeed also found in \cite{Wensink2012a}.  The active spinning
behavior is reminiscent of the rotational dynamics found in multicellular
spherical Volvox colonies \cite{Drescher2010}, but differs in the sense that
the glass phase lacks any orchestrated mechanism to direct the individual
particles' activity.

\subsection{Dynamics of the self-spinning active glass} 

\begin{figure}
        \begin{center}
    \includegraphics[width=0.48\textwidth]{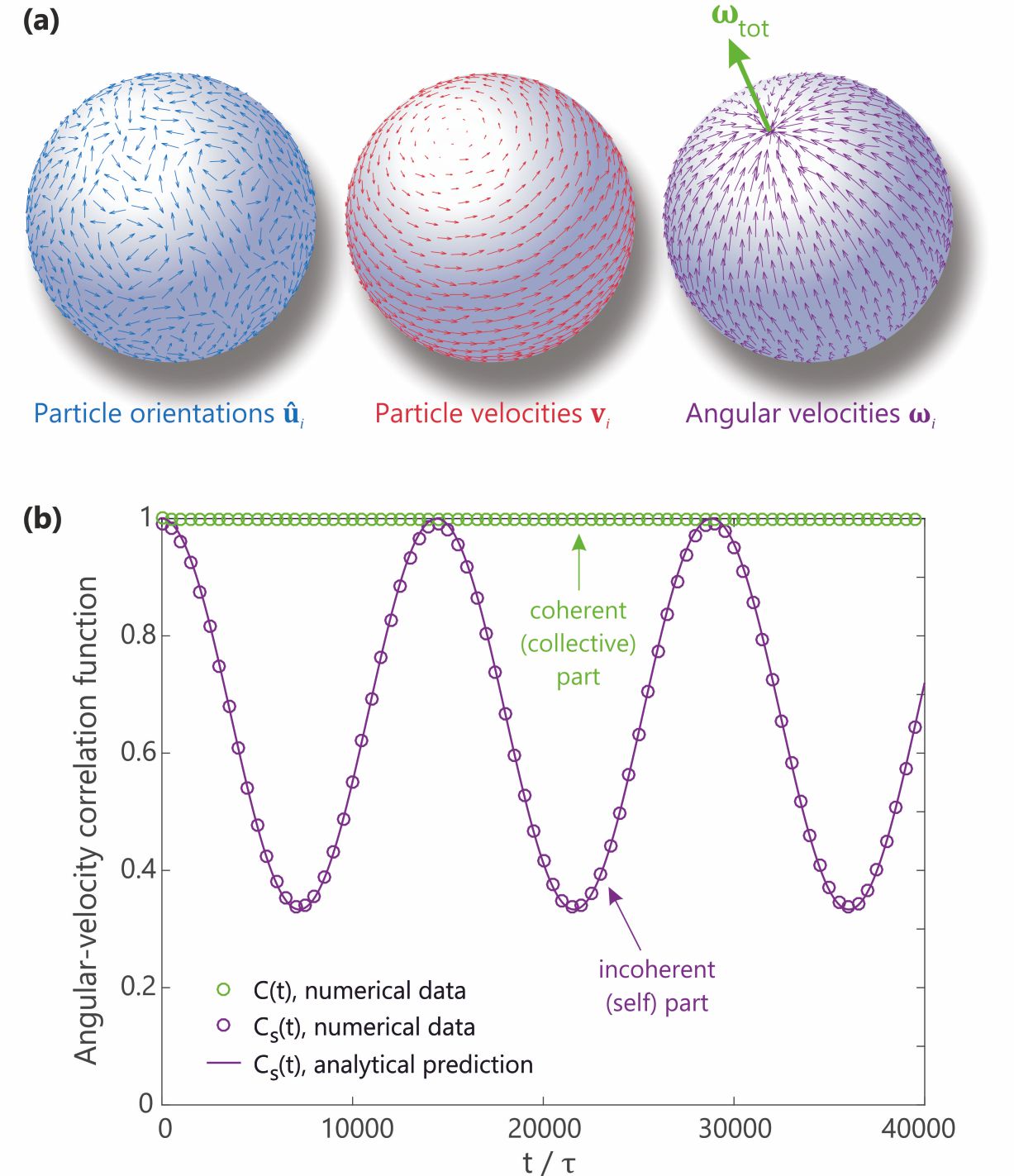}
  \end{center}
  \caption{
  \label{fig:angvel} 
\textbf{Angular velocities in the self-spinning glass phase.}
(a) Snapshots of the particle orientations $\hat{\mathbf{u}}_i$ (blue arrows),
instantaneous velocities $\mathbf{v}_i$ (red arrows), and normalized angular
velocities $\hat{\bm{\omega}}_i$ (purple arrows) for an arbitrary glassy
configuration of $N=800$ particles with aspect ratio $a=2$ at packing fraction
$\phi=0.5$. 
(b) Corresponding time correlation functions $C_{s}(t)$ and $C(t)$, probing the
self- and collective parts of the angular-velocity autocorrelation,
respectively. The solid purple line indicates the analytical prediction of
$C_s(t)$ for a rotation period of $T=14410\tau$.
  }
\end{figure}

In order to characterize the self-spinning motion, let us focus on the angular
velocity field in the glass phase. Figure \ref{fig:angvel}(a) depicts a
snapshot of the typical particle orientations $\hat{\mathbf{u}}_i$,
instantaneous velocities $\mathbf{v}_i$, and corresponding angular velocity
field for a glass of $N=800$ particles with aspect ratio $a=2$ and packing
fraction $\phi=0.5$, where the normalized angular velocity for each particle
$i$ is defined as $\hat{\bm{\omega}}_i = (\mathbf{r}_i \times
\mathbf{v}_i)/|\mathbf{r}_i||\mathbf{v}_i|$.  The total angular velocity,
defined as $\bm{\omega}_{\rm{tot}} = \sum_i \mathbf{r}_i \times \mathbf{v}_i$,
is a vector pointing in the direction of the rotation axis, whose norm
$|\bm{\omega}_{\rm{tot}}|$ quantifies the global angular speed of rotation.
The time-dependent dynamics of the spinning motion is now conveniently captured
in the autocorrelation function of the angular velocity. To this end, we make a
distinction between the incoherent or self-part of the correlation function
\begin{equation} 
C_s(t) = \frac{\sum_i \langle \hat{\bm{\omega}}_i(0) \cdot \hat{\bm{\omega}}_i(t) \rangle}
               {\sum_i \langle \hat{\bm{\omega}}_i(0) \cdot \hat{\bm{\omega}}_i(0) \rangle},
\end{equation}
and the coherent or collective part 
\begin{equation} 
C(t) = \frac{\langle \bm{\omega}_{\rm{tot}}(0) \cdot \bm{\omega}_{\rm{tot}}(t) \rangle}
             {\langle \bm{\omega}_{\rm{tot}}(0) \cdot \bm{\omega}_{\rm{tot}}(0) \rangle},
\end{equation}
where $t$ denotes time and the brackets are appropriate statistical averages.
As discussed below, these two functions offer valuable and complementary insight into 
the time-dependent dynamics of the system.  

Figure \ref{fig:angvel}(b) shows the time-dependent behavior of both
correlation functions, calculated for the glassy state depicted in Fig.\
\ref{fig:angvel}(a), where the statistical average is taken over different time
origins.  The incoherent function $C_s(t)$ clearly reveals a steady
non-vanishing rotational motion, with the period of rotation determined by the
net angular speed $|\bm{\omega}_{\rm{tot}}|$. We point out that this
oscillation period is essentially arbitrary; a different random starting
configuration will equilibrate to a different disordered state, giving rise to
a different net angular velocity. Indeed, we have performed tests for 1000
different initial conditions, and found that the Cartesian components of
$\bm{\omega}_{\rm{tot}}$ are normally distributed around zero, consistent with
the Central Limit Theorem.  Also note that $C_s(t)$ oscillates between the
values of 1 and 1/3, which is a consequence of the geometry of the spherical
surface: in the stable glass phase, the angular velocity of a particle at the
pole will anti-correlate with itself after half a period of rotation, while a
particle at the equator will have a constant angular velocity.  The total
particle average as a function of time, assuming homogeneous coverage of the
sphere, is then $C_s(t/T) = \int_0^1 dv [(2v-1)^2 \cos(2\pi t/T) - (2v-1)^2 +
1] = \frac{1}{3}[\cos(2\pi t/T) + 2]$, where $T$ is the total period of
rotation. As can be seen in Fig.\ \ref{fig:angvel}(b), this analytical result
is in perfect agreement with our numerical results.  For the \textit{coherent}
correlator $C(t)$, however, the curvature and topology of the confining
geometry do not play any role, since the total angular velocity
$\bm{\omega}_{\rm{tot}}$ is constant in the glassy state. Hence the normalized
collective autocorrelation function will always be 1 in this case. Overall,
these result confirm that the self-spinning glass state is a highly robust
phase that continues to spin indefinitely in an arbitrary but fixed direction.

\subsection{Melting and revitrification dynamics} 

\begin{figure*}
        \begin{center}
    \includegraphics[width=0.8\textwidth]{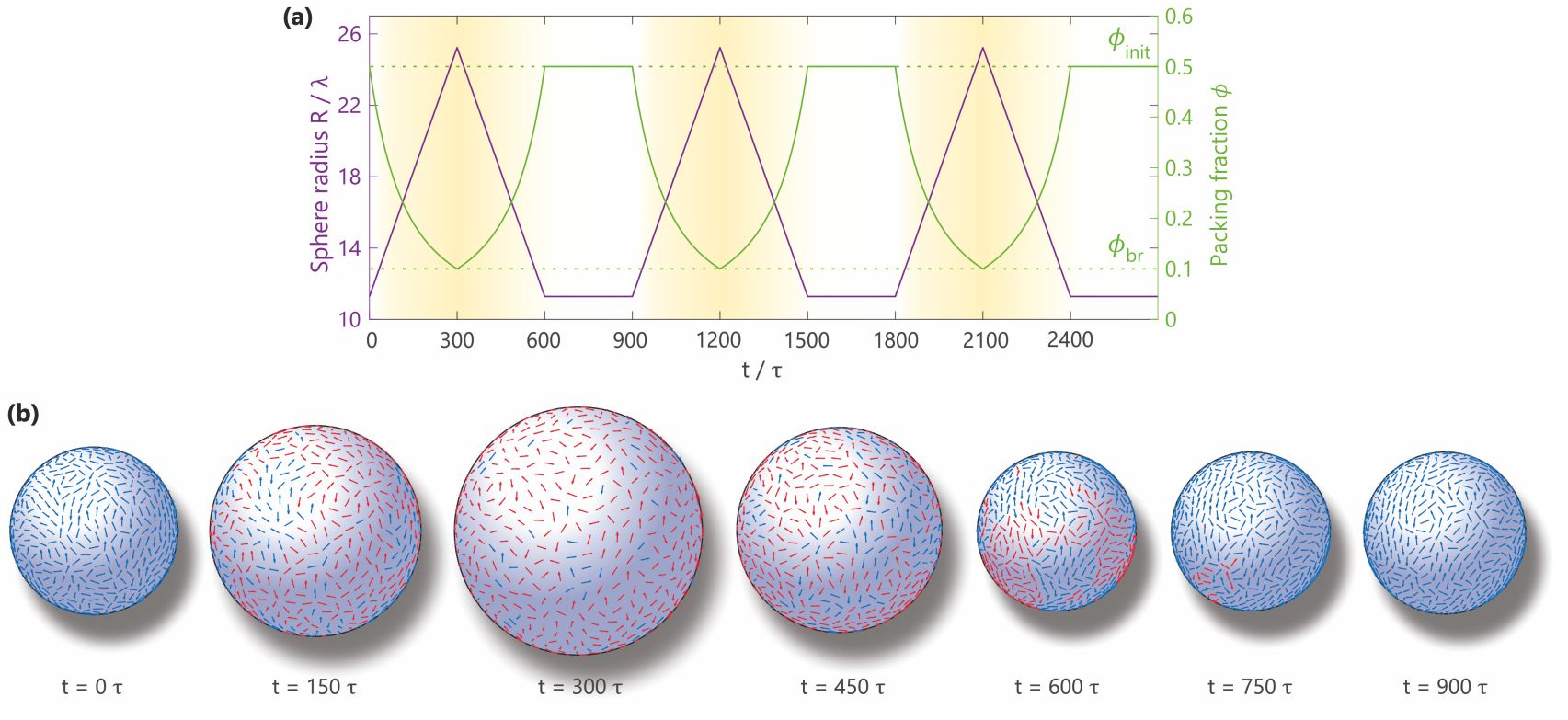}
  \end{center}
  \caption{
  \label{fig:breathing} 
  \textbf{Breathing protocol to induce melting and revitrification of the glass phase.}
(a) Protocol for three consecutive breathing cycles  of periodic inflation and
deflation of the sphere.  A single breathing cycle consists of three stages:
first we dilute the system from a packing fraction $\phi_{\rm{init}}$ to
$\phi_{\rm{br}}$ by linearly increasing the sphere radius $R$ in 30 steps,
allowing the system to briefly equilibrate at every new $R$-value for a
duration of $10\tau$.  We subsequently re-densify the system to the original
packing fraction $\phi_{\rm{init}}$ by a stepwise linear decrease in $R$ over a
time period of $30\times10\tau$, and finally we allow the system to
re-equilibrate at $\phi=\phi_{\rm{init}}$ during a time interval of $300\tau$.
In this example we have $\phi_{\rm{init}}=0.5$ and $\phi_{\rm{br}}=0.1$, as
indicated by the green dashed lines.
(b) Particle snapshots for a single breathing cycle with $\phi_{\rm{init}}=0.5$
and $\phi_{\rm{br}}=0.2$ (also see Supplementary Movies S6 and S7). Blue arrows indicate immobile particles whose centers
of mass have moved less than a distance $0.2\lambda$ within a time frame of
$20\tau$, while red arrows indicate mobile particles that have moved more that
$0.2\lambda$ during the same time interval. Notice that at low densities the
system melts and almost all particles undergo large displacements, while at
high densities the system locks into a glassy phase in which, aside from the
overall spinning motion, no particles rearrange.
  }
\end{figure*}

We now turn our attention to the dynamics that emerges upon melting and
revitrification of the spinning glass phase.  The relevant control parameter
that drives the glass transition in our system is the packing fraction, and
hence the active glass can be melted by increasing the size of the confining
sphere while keeping the particle number $N$ constant. The revitrification
process may subsequently be induced by compressing the sphere to a smaller
radius, thus effectively increasing the packing fraction again.  In order to
systematically study the effect of fluctuations in the packing fraction, we
introduce a ''breathing protocol" whereby the sphere is periodically inflated
and deflated to a certain upper and lower radius, respectively, allowing us to
switch repeatedly between the ergodic active-fluid phase and the dense glassy
state.  Figure \ref{fig:breathing}(a) illustrates the protocol for three
consecutive cycles that switch between packing fractions $\phi=0.5$ and
$\phi=0.1$, and Fig.\ \ref{fig:breathing}(b) shows typical snapshots of
particle configurations during one cycle (also see Supplementary Movies S6 and S7). 
In general, a single breathing cycle
starts at a packing fraction $\phi_{\rm{init}}$, and is then diluted to
$\phi_{\rm{br}} < \phi_{\rm{init}}$ by linearly increasing the sphere radius
$R$ in 30 steps. The system is subsequently re-densified toward
$\phi_{\rm{init}}$ by linearly decreasing $R$ in 30 steps, followed by a final
stage in which we keep the packing fraction constant at $\phi=\phi_{\rm{init}}$
(see Methods).  We note that this protocol is somewhat reminiscent of other
periodic driving schemes that are commonly applied to passive glasses, such as
oscillatory shearing \cite{Hyun2011} and thermal cycling
\cite{Zhao2013,Ketov2015}.  However, our breathing protocol amounts to a
periodic change in density, while shearing and thermal cycling keep the density
constant.

\begin{figure}
        \begin{center}
    \includegraphics[width=0.47\textwidth]{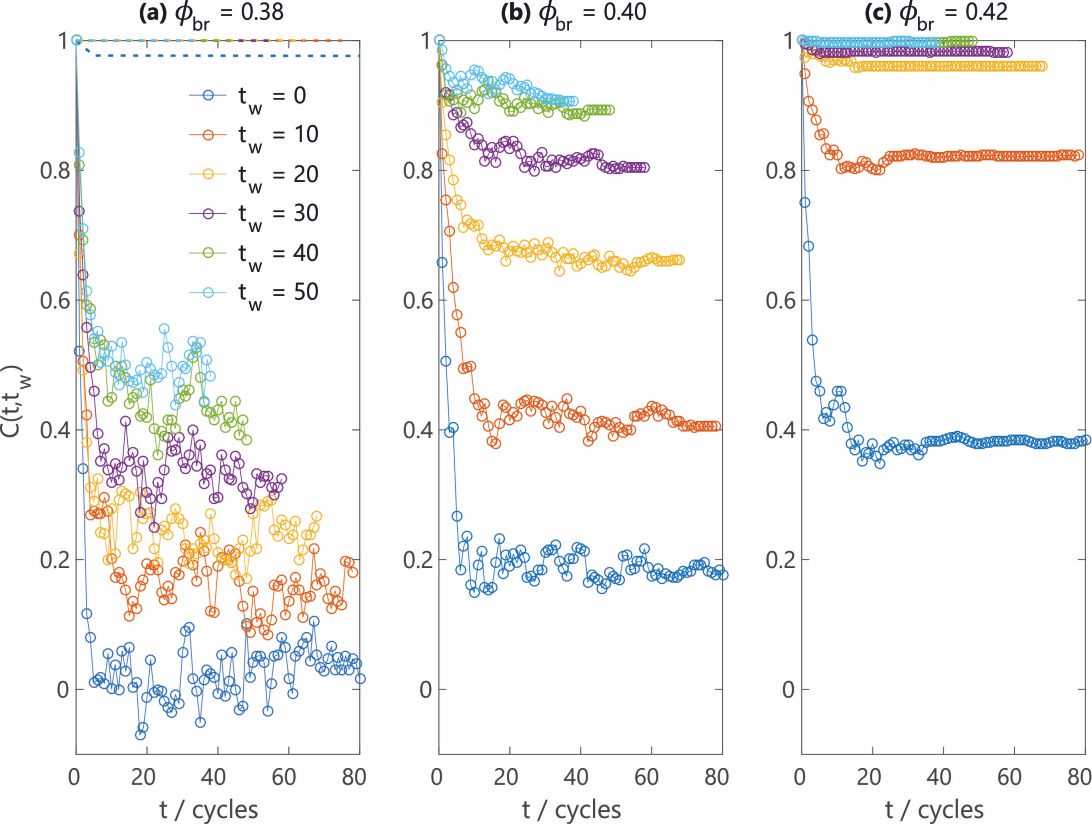}
  \end{center}
  \caption{
  \label{fig:Ctw} 
\textbf{Aging dynamics in the active glass phase upon melting and re-vitrification.}
Time correlation function of the total angular velocity, $C(t,t_w)$, for
different waiting times $t_w$ and different breathing amplitudes
$\phi_{\rm{br}}$: (a) $\phi_{\rm{br}}=0.38$, (b) $\phi_{\rm{br}}=0.40$, and (c)
$\phi_{\rm{br}}=0.42$, all starting from the active glass phase at
$\phi_{\rm{init}}=0.5$. The data were collected for $N=400$ particles with
$a=2$, averaged over 100 independent equilibrated starting configurations. As a
reference, we also show the results for a \textit{passive} system without any
self-propulsion, plotted as dashed lines in panel (a) for
$\phi_{\rm{br}}=0.38$.
  }
\end{figure}

Let us investigate the dynamics of the system as a function of the number of
applied breathing cycles.  To this end, we introduce a time-dependent
angular-velocity correlation function
\begin{equation} 
C(t,t_w) = \frac{\langle \hat{\bm{\omega}}_{\rm{tot}}(t_w) \cdot \hat{\bm{\omega}}_{\rm{tot}}(t+t_w) \rangle}
                {\langle \hat{\bm{\omega}}_{\rm{tot}}(t_w) \cdot \hat{\bm{\omega}}_{\rm{tot}}(t_w) \rangle},
\end{equation}
which now depends explicitly on the waiting time $t_w$. Here,
$\hat{\bm{\omega}}_{\rm{tot}}$ is the normalized total angular velocity and the
brackets denote an average over different independent configurations (also see
Methods).  For convenience, we will quantify $t_w$ in units of the applied
number of cycles, with each cycle representing a time span of $900\tau$.
Interestingly, we find a distinctly different behavior of $C(t,t_w)$ depending
on the magnitude of the fluctuations in the packing fraction.  Figure
\ref{fig:Ctw} compares the dynamics in a system of $N=400$ particles ($a=2$)
for different breathing amplitudes of $\phi_{\rm{br}}=0.38, 0.40$, and 0.42,
all starting from a glassy phase at $\phi_{\rm{init}}=0.5$.  For the largest
expansion amplitude considered, $\phi_{\rm{br}}=0.38$, it can be seen that
$C(t,t_w)$ rapidly decays to zero if $t_w=0$ (i.e., before applying any
expansion-compression cycle), but builds up an increasingly large nonzero
long-time limit as the number of applied cycles increases. This signifies that
the total angular velocity in the glassy state becomes increasingly more
correlated to that of all future revitrified configurations.  Applying a
breathing protocol with a slightly smaller change in density, e.g.\
$\phi_{\rm{br}}=0.40$, enhances this effect.  In view of this marked
dependence on waiting time, which we observe both in $N=400$ and $N=800$
systems under moderately small breathing amplitudes, we assert that our system
is \textit{aging}.

It is important to realize that the particles' inherent activity is a crucial
ingredient for the aging process; an equilibrated \textit{passive} system
without self-propulsion will--in the absence of noise--remain in the same
configuration indefinitely, regardless of breathing amplitude and waiting time
$t_w$. Indeed, as shown for comparison for $\phi_{\rm{br}}=0.38$ in Fig.\
\ref{fig:Ctw}(a), a strictly passive reference system rapidly yields a constant
correlation value $C(t,t_w)=1$, exhibiting only a marginal decorrelation effect
at very short times (also see Supplementary Movie S8). This stark contrast between the active and passive
time-dependent dynamics confirms that the observed aging phenomenon is indeed
activity-induced. 

The degree of aging is, however, sensitive to the relative amplitude of the breathing motion. 
Under the mildest
breathing protocol considered here, $\phi_{\rm{br}}=0.42$, the correlator
$C(t,t_w)$ already attains its maximal value of 1 after 40 full cycles, and hence
no more aging dynamics can be observed for all longer waiting times $t_w>40$.
In the extreme limit of $\phi_{\rm{br}} = \phi_{\rm{init}}$, the system remains
frozen in its original spinning-glass configuration for all times $t$ and $t_w$, 
thus causing the aging effect to vanish completely. As a final case, let us consider the 
opposite limit of $\phi_{\rm{br}} \rightarrow 0$, which allows the system to melt into 
a dilute fluid phase during every cycle (also see Supplementary movie S6). For such large-amplitude breathing, $C(t,t_w)$ 
will rapidly decay to zero, independent of the number of cycles $t_w$. That is,
the re-solidification stage from the melt at $\phi_{\rm{br}}$ to $\phi_{\rm{init}}$ will always
yield a new glassy configuration that is completely uncorrelated to the orientation of
$\bm{\omega}_{\rm{tot}}$ at the beginning of the cycle. In analogy to
the phenomenology in oscillatory-sheared passive systems \cite{Viasnoff2002},
we will refer to 
such a process as \textit{rejuvenation}: each full breathing cycle will wash
away any possible memory of the original glassy state and produce a new
self-spinning glass with an entirely new $\bm{\omega}_{\rm{tot}}$. 

\subsection{Activity-induced aging mechanism}

We now seek to gain more insight into the physical mechanism that underlies the
observed activity-induced aging dynamics. Upon inspection of the particle
trajectories for $\phi_{\rm{init}}=0.5$ and $\phi_{\rm{br}}\approx 0.4$, we
find that the system generally does not melt completely, but rather exhibits a
limited amount of cooperative particle rearrangements--strongly reminiscent of
the dynamically heterogeneous dynamics observed in normal glass-forming
liquids. As the aging process further evolves, the average number of
rearranging particles tends to decrease and ultimately the system locks into a
new configuration in which all relative particle motion has ceased.  That is,
the system has seemingly reached a glassy state that is sufficiently stable to
sustain a breathing amplitude of $\phi_{\rm{br}}$, and consequently remains in
this stable state indefinitely (see Supplementary Movie S7).

\begin{figure}
        \begin{center}
    \includegraphics[width=0.5\textwidth]{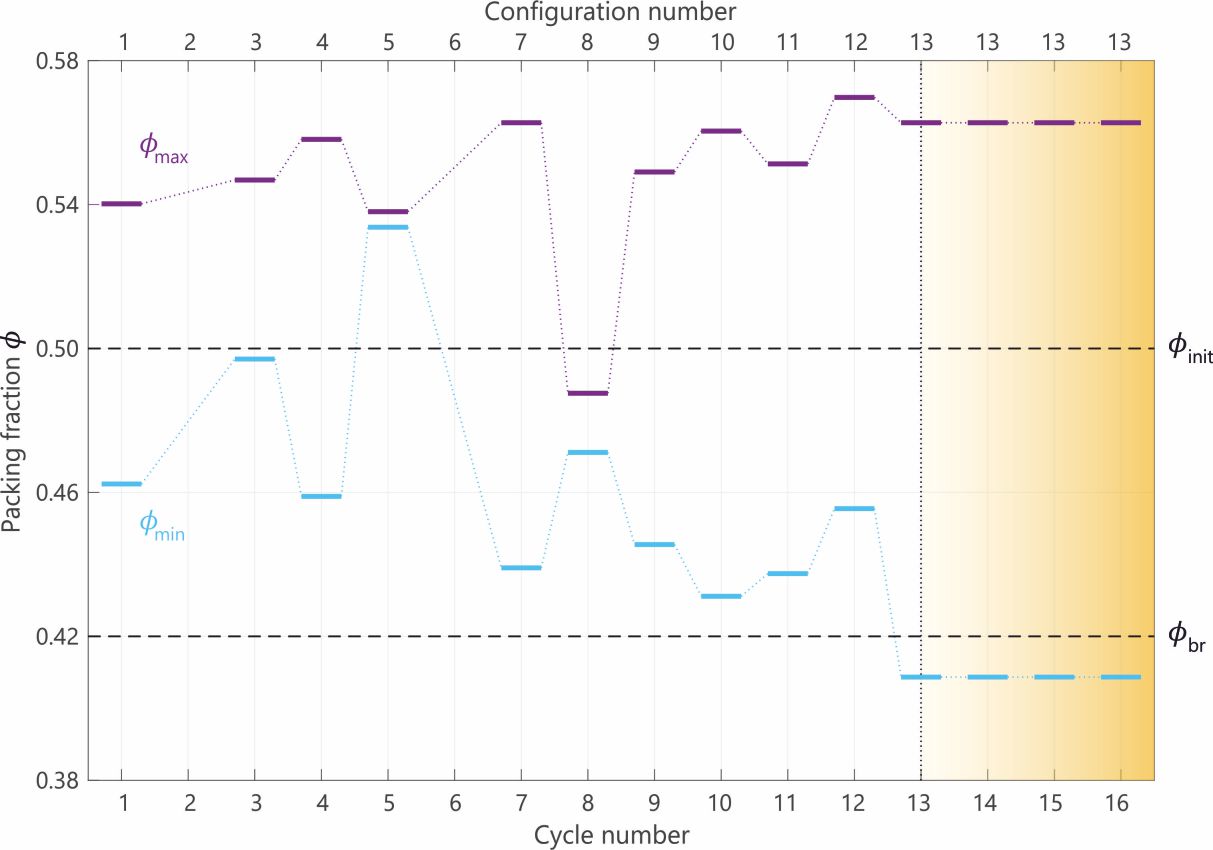}
  \end{center}
  \caption{
  \label{fig:stabtraj} 
  \textbf{Stability of the active-particle configurations formed during aging.}
  Stability analysis of the configurations obtained from a single trajectory
for $N=800$ active particles with self-propulsion strength $F=1$, undergoing 20
consecutive breathing cycles for $\phi_{\rm{init}} = 0.5$ and $\phi_{br} =
0.42$.  The stability was measured for the final configuration after every
cycle. The purple lines indicate the maximum packing fraction, $\phi_{\rm{max}}$,
at which a configuration is still stable; for $\phi > \phi_{\rm{max}}$ the
system will undergoing particle rearrangements to avoid unphysical particle
overlaps. The blue lines indicate the minimum packing fraction,
$\phi_{\rm{min}}$, at which a configuration is still stable; for $\phi <
\phi_{\rm{min}}$ the system will melt into an active fluid phase. Note that
after 13 cycles, the system locks into a configuration that is stable at all
$\phi_{\rm{br}} < \phi < \phi_{\rm{init}}$, and hence remains in this
configuration for all remaining cycles (shaded yellow region).
  }
\end{figure}

In order to quantify the emergent stability of the particle configurations
during aging, we use the total number of rearranging particles $N_r$ as a
metric and determine at which packing fraction a given configuration will
become unstable such that $N_r > 0$. Here we define particle rearrangement
using a Lindemann-like criterion for melting, as described in the Methods
section.  Figure \ref{fig:stabtraj} shows the results of our stability analysis
for a single aging trajectory of $N=800$ active particles undergoing 16
consecutive breathing cycles between $\phi_{\rm{init}}=0.5$ and $\phi_{\rm{br}}
= 0.42$. The stability was measured for every configuration at the end of a
full cycle.  Let us first point out two general observations with respect to
Fig.\ \ref{fig:stabtraj}: first of all, there need not exist any value of
$\phi$ for which a given configuration is stable. Indeed, configuration numbers
2 and 6 in Fig.\ \ref{fig:stabtraj} are unstable for all possible packing
fractions. Secondly, if there exists a range of $\phi$ values for which an
active configuration is stable, the stability range will be bounded both from
above and from below. Figure \ref{fig:stabtraj} shows these upper and lower
packing fractions--denoted as $\phi_{\rm{max}}$ and $\phi_{\rm{min}}$,
respectively--for all the remaining configurations of the trajectory. The
reason for this boundedness is as follows: at high packing fractions $\phi >
\phi_{\rm{max}}$, the particles are forced to rearrange in order to avoid
unphysical overlaps due to the short-range repulsive interactions.  For low
packing fractions $\phi < \phi_{\rm{min}}$, the distance between particles
becomes sufficiently large to facilitate quasi-ergodic particle motion, causing
the system to ultimately melt into an active fluid phase. It is important to
note that the latter lower bound does not exist for passive systems: particles
with zero self-propulsion will become completely immobile ($N_r=0$) in the
limit of $\phi \rightarrow 0$, thus rendering them strictly stable in our
definition.  We will expand upon this point in the Discussion section. 

Let us now return to the activity-induced aging phenomenon. Fig.\ \ref{fig:stabtraj} reveals that
the stability of the configurations formed during a single breathing trajectory
does not monotonically increase with time; on the contrary, we observe an erratically
varying pattern of stabilities, 
including intermittent
states (after cycle 2 and 6) that are strictly unstable for all $\phi$. After
13 cycles, however, the system reaches a configuration whose stability range
spans the \textit{entire} amplitude of the breathing motion, i.e.,
$\phi_{\rm{max}} > \phi_{\rm{init}}$ and $\phi_{\rm{min}} < \phi_{\rm{br}}$.
Once this stable state is reached, the applied breathing protocol can no longer
destabilize the configuration, consequently prohibiting the system to explore
any other configurations in the remaining cycles.  In close analogy to work on
periodically driven systems \cite{Corte2008}, we thus conclude that our system
has undergone an irreversible, random self-organization process toward an
''absorbing state" in which all particle fluctuations have vanished. It is this
mechanism that underlies the observed aging: the system continues to explore
many different configurations until it spontaneously reaches a stable state
from which it can no longer escape.

We can define the absorbing state more generally as the firstly formed
configuration with stability bounds $\phi_{\rm{max}} \geq \phi_{\rm{init}}$ and
$\phi_{\rm{min}} \leq \phi_{\rm{br}}$; note that this state is principally one
of infinitely many possible configurations.  For moderately breathing
amplitudes, such a stable state may always be reached provided that the waiting
time is sufficiently long, as can be seen from Fig.\ \ref{fig:Ctw}.
Conversely, for a strictly passive system with zero self-propulsion,
\textit{any} configuration in which the particles do not overlap too strongly
can act as an absorbing state, and hence we observe no notable aging dynamics
in the passive case.

\begin{figure}
        \begin{center}
    \includegraphics[width=0.5\textwidth]{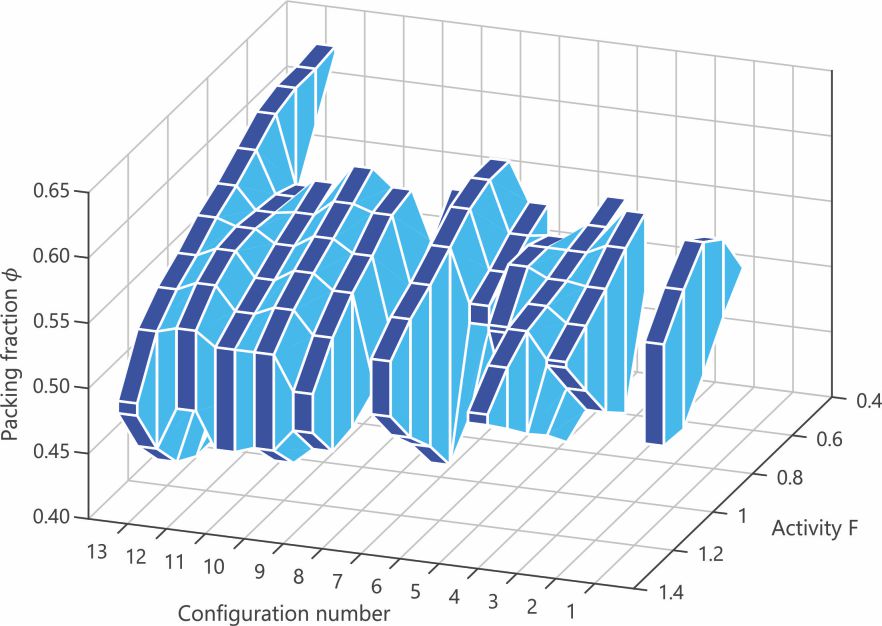}
  \end{center}
  \caption{
  \label{fig:stab3D} 
  \textbf{Stability of active configations as a function of self-propulsion strength.} 
  Stability analysis for the same configurations as in Fig.\ \ref{fig:stabtraj}, 
  but for different activities $F$. 
  Dark-blue shaded areas enclose the regions of stability ($\phi_{\rm{min}} \leq \phi \leq \phi_{\rm{max}}$).
  Note that in this specific example, configuration number 2 
  remains unstable for all possible values of $F>0$ considered; such a configuration can only be stabilized
  for $F=0$ in the non-interacting gas limit $\phi \rightarrow 0$. Configuration number 6, which
  is unstable for $F=1$, becomes stable for $0.6 \leq F \leq 0.9$.
  }
\end{figure}

Finally, let us investigate how the stability of the formed configurations--and
thus the nature of the absorbing state--is affected by the magnitude of the
self-propulsion force $F$.  As a proof-of-principle study, we have measured the
stability of the 13 unique particle configurations considered in Fig.\
\ref{fig:stabtraj} for different values of $F$, thereby keeping the initial
particle positions and orientations the same as for the $F=1$ reference case
(see Methods). The results are shown in Fig.\ \ref{fig:stab3D}. Clearly, the
stability dependence for a given configuration on $F$ is highly non-monotonic:
the upper ($\phi_{\rm{max}}$) and lower ($\phi_{\rm{min}}$) stability bounds
can both increase or decrease with increasing $F$, and also the total width of
the stability range, i.e.\ $\phi_{\rm{max}}-\phi_{\rm{min}}$, depends strongly
on the exact configuration and value of $F$.  In view of these results, we
conclude that the set of possible absorbing states will generally be different
for different values of the self-propulsion strength. This may also be
rationalized by considering that the stability in our active glassy system
arises from a delicate balance between the intrinsic self-propulsion and
repulsive pair-interaction forces on the particles; changing the magnitude of
the active forces will generally alter the force balance across the entire
disordered network, giving rise to either reduced or enhanced local stability
in the system. Consequently, the first absorbing state that a system finds is
sensitively dependent on the exact value of $F$.

\section{Discussion} 
\begin{figure}
        \begin{center}
    \includegraphics[width=0.45\textwidth]{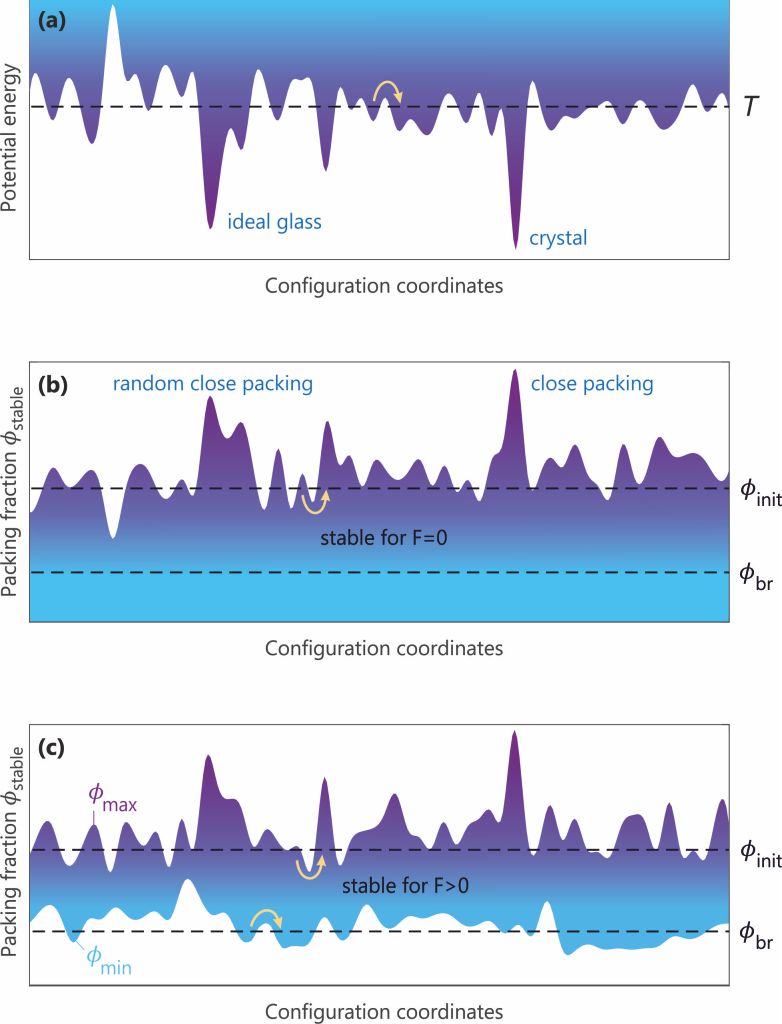}
  \end{center}
  \caption{
  \label{fig:landscape} 
\textbf{Schematic illustration of the landscape picture in glassy physics.}
The $x$-axis represents all configurational coordinates of an $N$-particle system.
(a) The traditional potential-energy landscape of passive glass-forming
systems, adapted from Ref.\ \cite{Debenedetti2001}, with a typical temperature
$T$ indicated by the dashed line.  The global minimum of the energy is assumed
to be the crystalline state, while the lowest energy state for a disordered
configuration is the ideal glass.
(b) Schematic stability landscape of passive ($F=0$) glass-formers.  The
blue-shaded region marks the range of packing fractions for which the different
configurations are stable, with stability defined here in terms of a
Lindemann-like melting criterion.  The global stability maximum is, by
definition, the close-packing configuration, and for disordered systems the
maximum corresponds to random close packing.
(c) Schematic stability landscape of active ($F>0$) glass-forming systems.
Every active configuration will generally melt at sufficiently low density, and
consequently the lower stability bound $\phi_{\rm{min}}$ must be larger than
zero. The dashed lines in panels (b) and (c) indicate the range of packing
fractions, $\phi_{\rm{br}} \leq \phi \leq \phi_{\rm{init}}$, in which we
prepare the system.  The yellow arrows indicate typical barrier-crossing events
to different parts of configuration space.
  }
\end{figure}

As discussed earlier, the observed aging dynamics occurs only for inherently
active systems with a nonzero self-propulsion strength $F>0$. Let us now
compare this novel activity-induced aging mechanism with conventional aging in
passive glass-forming systems. A key paradigm in the phenomenology of
non-active glasses is the potential energy landscape
\cite{Stillinger1995,Debenedetti2001,Kob2000}--a generally highly complex and rugged
surface that describes the total potential energy of the system as a function
of the $3N$-dimensional configuration space [see Fig.\ \ref{fig:landscape}(a)].
Within this landscape picture, aging and rejuvenation are understood as
out-of-equilibrium processes whereby the system visits deeper or shallower
local energy minima, respectively. The energy barriers separating these minima
may be surmounted due to thermal fluctuations; if the system is prepared at a
temperature $T$, the typical barrier height that can be crossed is on the order
of $k_{\rm{B}}T$, with $k_{\rm{B}}$ denoting the Boltzmann constant. This
passive energy-landscape scenario is illustrated schematically in Fig.\
\ref{fig:landscape}(a).  Note that here the global energy minimum corresponds
to the crystalline state, and the lowest minimum for a disordered configuration
is referred to as the ideal glass state.  For inherently active systems,
however, the total potential energy is not neccessarily a useful metric, since
the self-propulsion of the particles requires a constant (implicit) source of
energy. Indeed, we also find that the total potential energy of our active
system is generally not minimized during aging, implying that the aging process
in passive glasses is not equivalent to our active-matter case.

Instead, we argue that the observed active aging and rejuvenation dynamics can
be associated with a rugged ''stability landscape" that quantifies the
mechanical stability  of all possible particle configurations. Such a landscape
is essentially the $3N$-dimensional generalization of Fig.\ \ref{fig:stabtraj}
discussed in the previous section, where we have defined stability in terms of
a Lindemann-like melting criterion.  Let us first consider the passive version
of this landscape. Since a configuration with $F=0$ (in the absence of noise)
will always be stable such that $N_r=0$ in the dilute limit $\phi \rightarrow
0$, we have a rigorous minimum stability bound $\phi_{\rm{min}}=0$ and a
maximum bound $\phi_{\rm{max}}$ that depends on the exact configuration.
Figure \ref{fig:landscape}(b) shows a schematic representation of this passive
scenario.  Note that the global maximum of $\phi_{\rm{max}}$ is, by definition,
the close-packing configuration, and for \textit{disordered} systems the
maximum attainable value of $\phi_{\rm{max}}$ is at random close-packing.

For an active system with $F>0$, however, the shape and properties of the
stability landscape become decidedly different: first of all, all active
configurations must melt at sufficiently low density, so that
$\phi_{\rm{min}}>0$.  Moreover, activity may both enhance and reduce the
stability of a given configuration, and hence the positions of local minima and
maxima will generally shift with varying $F$.  It must also be noted that
certain configurations can become strictly unstable for all $F>0$, as we found
for e.g.\ configuration number 2 in Figs.\ \ref{fig:stabtraj} and
\ref{fig:stab3D}, giving rise to open voids in the stability landscape;
however, in analogy to the inherent-structure formalism for passive glasses
\cite{Stillinger1982,Kob2000,Debenedetti2001}, we assume that any instantaneous
configuration can be quenched to a nearby state with a finite stability range,
and the landscape of such inherent structures will be devoid of voids.  With
this active-stability picture in mind, as illustrated in Fig.\
\ref{fig:landscape}(c), we can interpret the aging process as follows: during a
dynamics simulation with periodic breathing, the system will explore different
regions of the landscape until it reaches an absorbing state that is
characterized by $\phi_{\rm{max}} \geq \phi_{\rm{init}}$ and $\phi_{\rm{min}}
\leq \phi_{\rm{br}}$. Rejuvenation can occur by subsequently increasing the
amplitude of breathing to $\phi_{\rm{br}} < \phi_{\rm{min}}$, inducing a
(partial) melt to allow the system to explore new regions of configuration
space, until eventually a new absorbing state is reached with higher stability.
We emphasize that this aging and rejuvenation analysis should apply generally
to \textit{any} active glassy system, regardless of the system size and
topology, and is thus not limited to the spherical active-rod model of the
present study.

As a final point, let us elaborate on the role of noise in the observed active
aging dynamics.  In the noise-free case, the aging process ceases as soon as
the active system reaches an absorbing state; however, if noise is added by
introducing fluctuations $\Delta \phi$ in the breathing amplitude, the system
might be able to escape from an absorbing state and cross local barriers on the
stability landscape whose heights are on the order of $\Delta \phi$. Such
fluctuations would essentially play the role of thermal fluctuations in the
passive case, and would cause the active aging process to continue
indefinitely.  Indeed, just as a passive thermal glass will age by visiting
increasingly deeper energy minima, our active glass is expected to reach
increasingly more stable states as it ages under a weakly fluctuating breathing
motion.  This barrier-crossing process is illustrated schematically in Fig.\
\ref{fig:landscape}; note that for active glasses a stability barrier may exist
both in $\phi_{\rm{max}}$ and $\phi_{\rm{min}}$. In addition to this source of
fluctuations, we may also consider \textit{thermal} noise in our system, which
can give rise to stochastic fluctuations in the particles' centers of mass and
orientations.  In such a case, we expect every absorbing state to be replaced
by a \textit{basin} of absorbing states, analogous to the passive
potential-energy landscape picture where basins emerge as deep energy minima
that are separated by relatively small barriers.  Importantly, however, within
the current stability-landscape picture, thermal noise can also act as a proxy
for activity: a passive particle system will, in the presence of thermal
fluctuations, melt at sufficiently low densities.  Hence, the lower stability
bound $\phi_{\rm{min}}$ will always become greater than zero, akin to the
noise-free case of Fig.\ \ref{fig:landscape}(c) for active systems.  Finally,
we note that the existence of noise may also provide opportunities for encoding
memory into an active system, similar to recent studies on passive model
glass-formers under oscillatory shear \cite{Fiocco2014}.

In conclusion, we have explored the emergent dynamics in an active-matter
system constrained to a spherical manifold. In the absence of strong aligning
forces, we find that active particles at sufficiently high density can undergo
a glass transition towards a non-ergodic state that is characterized by
persistent collective spinning motion. Upon repeated melting and
revitrification of such a self-spinning glass, we observe signatures of
non-equilibrium aging and rejuvenation that occur solely for strictly active
systems. We rationalize the activity-induced aging process in terms of a
mechanical stability landscape: as the active system ages, it randomly explores
different regions of configuration space until it reaches an absorbing state
that is sufficiently stable to resist melting.  We expect our results to hold
generally for active systems that can form a glassy phase, regardless of system
size and topology.  Our findings may be experimentally verified in e.g.\ dense
suspensions of biological or artificial microswimmers confined to a liquid
droplet interface or hydrogel.

\section{Methods} 
{\footnotesize 

\subsection{Model system and dynamics simulations}
Our active-matter system is composed of $N$ interacting rods of length $\ell$
that all experience a constant self-propulsion force with magnitude $F$ along
their longitudinal rod axis $\hat{\mathbf{u}}$. In order to mimic steric
repulsion between the particles, we represent each rod $i$ as a rigid chain of
$n$ spherical segments ($n=\lceil14\ell/8\rceil$), and let every segment
interact with all the segments of any other rod $j$ through a repulsive Yukawa
potential. The total interaction energy between a pair of rods is given by
$U_{ij} = \frac{U_0}{n^2} \sum_{\alpha=1}^n \sum_{\beta=1}^n
\frac{\exp(-r_{ij,\alpha\beta}/\lambda)}{r_{ij,\alpha\beta}}$, where
$r_{ij,\alpha\beta}$ is the Euclidean distance between segment $\alpha$ of rod
$i$ and segment $\beta$ of rod $j$, $U_0$ is the strength of the potential, and
the screening length $\lambda$ can be interpreted as the effective diameter of
the segments.  Note that in terms of computational costs, our force-calculation
routine is effectively that of an ($N\times n$)-particle system, rather than
$N$.

We simulate the active-particle dynamics by integrating the overdamped Brownian
equations of motion for the center-of-mass coordinates $\mathbf{r}_i$ and
normalized orientation vector $\hat{\mathbf{u}}_i$ of each particle $i$.
Explicitly, we consider the dynamics within the \textit{local 2D plane
tangential to the sphere} at position $\mathbf{r}_i$, project all segment
coordinates and $\hat{\mathbf{u}}_i$ onto this plane, and solve
\begin{eqnarray}
  \dot{\mathbf{r}}_i &=& \mathbf{D}_T [-\nabla_{\mathbf{r}_i} U + F \hat{\mathbf{u}}_i], \nonumber \\ 
\label{eq:BD}
  \dot{\hat{\mathbf{u}}}_i &=& -\mathbf{D}_R \nabla_{\hat{\mathbf{u}}_i} U,
\end{eqnarray}
where the dots denote time derivatives, $U = (1/2) \sum_{i, j\neq i} U_{ij}$,
and $\nabla_{\hat{\mathbf{u}}_i}$ is the gradient on the unit circle.
The matrices $\mathbf{D}_T$ and $\mathbf{D}_R$ represent inverse translational and rotational
friction tensors, respectively, defined as
\begin{eqnarray}
  \mathbf{D}_T &=& D_0 [D_{\parallel}\hat{\mathbf{u}}_i \otimes \hat{\mathbf{u}}_i 
                  + D_{\perp}(\mathbf{I} - \hat{\mathbf{u}}_i \otimes \hat{\mathbf{u}}_i)], \\
  \mathbf{D}_R &=& D_0 D_R \mathbf{I},
\end{eqnarray}
where $D_0$ is the Stokesian diffusion coefficient, 
$\mathbf{I}$ is the $2\times2$ unit matrix, $\otimes$ is the dyadic product, and 
for the parameters $D_{\parallel}$, $D_{\perp}$, and $D_R$ we use, as in Ref.\ \cite{Wensink2012,Wensink2012a},
the standard expressions for rod-like macromolecules given in Ref.\ \cite{Tirado1984},
\begin{eqnarray}
 2\pi D_{\parallel} &=& \ln(a) - 0.207 + 0.980 a^{-1} - 0.133a^{-2}, \nonumber \\
 4\pi D_{\perp} &=& \ln(a) + 0.839 + 0.185a^{-1} + 0.233a^{-2}, \nonumber \\
 \pi a^2 D_R/3 &=& \ln(a) - 0.662 + 0.917a^{-1} - 0.050 a^{-2}.
\end{eqnarray}
After every time step in the propagation of Eq.\ (\ref{eq:BD}), we project the
coordinates and orientation vector $\hat{\mathbf{u}}_i$ onto the tangent plane
at the particle's new position $\mathbf{r}_i$.  Finally, we note that the
equations of motion (\ref{eq:BD}) do not contain any stochastic terms, implying
that the dynamics is fully deterministic and is governed solely by the
repulsive pair interactions and self-propulsion forces.

Following Ref.\ \cite{Wensink2012a}, we adopt characteristic units such that
$\lambda=1$, $F=1$, and $D_0=1$, implying that time is measured in units of
$\tau = \lambda/(D_0 F)$.  We fix the strength of the interaction potential to
$U_0=250$ and include only segment-segment interactions that fall within a
cutoff radius $r_c=6\lambda$.  For the remaining parameters in our simulations,
namely the total particle number $N$, the rod aspect ratio $a$, and packing
fraction $\phi$, we typically use values of $N=400$ or 800, $1.5 \leq a \leq
16$, and $0.01 \leq \phi \leq 0.7$.  All simulations are performed using an
Euler integration scheme with a discrete time step of $0.01\tau$.

Independent starting configurations are produced by first placing all
particles' centers of mass randomly on a spherical surface with large radius
$R_0 \geq \sqrt{N \ell \lambda / (0.4\pi)}$ (corresponding to dilute packing
fractions $\phi \leq 0.1$), using spherical particle coordinates $\mathbf{r}_i
\equiv (r,\theta_i,\varphi_i) = (R_0, \cos^{-1}(2x_1-1), 2\pi x_2)$, where
$\theta_i$ and $\varphi_i$ are the polar and azimuthal angles of particle $i$,
respectively. The variables $x_1$ and $x_2$ are drawn randomly from a uniform
distribution on the interval $(0,1)$ to ensure approximately uniform coverage
on the spherical surface.  Similarly, we generate random particle orientations
on the unit sphere, $\hat{\mathbf{u}}_i = (1, \cos^{-1}(2x_3-1), 2\pi x_4)$,
where again $x_3$ and $x_4$ are random variates on $(0,1)$.  We subsequently
project these orientation vectors onto the local tangent plane at position
$\mathbf{r}_i$ and normalize such that $|\hat{\mathbf{u}}_i|=1$. In order to
remove any unphysical overlaps between rods, we randomly displace particles
whose segment coordinates overlap to within a distance of $\lambda$.  After
generating such an overlap-free random configuration at very low density, we
linearly decrease the sphere radius from $R_0$ to the desired size $R$
(corresponding to the desired packing fraction $\phi_{\rm{init}}$) in 200
steps, thereby allowing the system to briefly equilibrate for a time duration
of $1\,\tau$ at every fixed radius.  We then let the system equilibrate at
$\phi = \phi_{\rm{init}}$ for a duration of $2000\tau$, and subsequently
collect data for analysis over a period of $60\,000\tau$.

\subsection{Melting and revitrification dynamics protocol}
A single breathing cycle starts at a packing fraction $\phi_{\rm{init}}$, and
is then diluted to $\phi_{\rm{br}} < \phi_{\rm{init}}$ by linearly increasing
the sphere radius $R$ in 30 steps, allowing the system to briefly equilibrate
at every new $R$-value for a duration of 10$\tau$.  The system is subsequently
re-densified toward $\phi_{\rm{init}}$ by linearly decreasing $R$ again over
$30\times10\tau$, followed by a final stage in which we keep the packing
fraction constant at $\phi=\phi_{\rm{init}}$ during $300\tau$. Note that
the time it would take a single free rod of length $\ell=2\lambda$ to swim its
own length is $13.32\tau$, and the total cycle period thus offers a reasonable
compromise between a quasi-static and sudden quench.

The autocorrelation functions of the angular velocity are calculated based on
the angular velocities in the final configuration of every full breathing
cycle.  For the passive ($F=0$) reference case for $\phi_{\rm{br}}=0.38$ [Fig.\
\ref{fig:Ctw}(a)], we find that all the instantaneous velocities
${\mathbf{v}}_i$ are virtually zero, thus obscuring the calculation of the
angular velocities $\hat{\bm{\omega}}_i = (\mathbf{r}_i \times
\mathbf{v}_i)/|\mathbf{r}_i||\mathbf{v}_i|$ with large numerical noise. In
order to still probe any possible changes in the passive particle
configuration, we have assumed $\mathbf{v}_i=\hat{\mathbf{u}}_i$ in this case.
As can be seen from the dashed lines in Fig.\ \ref{fig:Ctw}(a), we detect only
very small displacements for passive particles (leading to a decorrelation of
$C(t,t_w)$ from 1 to $\approx 0.97$), and only at very short initial times
($t<5$ cycles).  Note that these marginal rearrangements are essentially a
consequence of the softness of the pair interaction; if the particles would
interact through a strictly hard potential, an overlap-free configuration
would--in the absence of activity and noise--rigorously yield $C(t,t_w)=1$.

\subsection{Stability analysis}
In order to quantify the stability of the particle configurations during aging,
we use the total number of displaced particles $N_r$ as a metric. More
specifically, for a given aging trajectory, we first place every configuration
that is formed after a full breathing cycle onto a new sphere of varying radius
$R_s$ ($R_1 > R_s > R_2$), where $R_s$ is varied linearly in 500 steps from
$R_1$ to $R_2$. We choose these upper and lower bounds of the sphere radius
such that they correspond to packing fractions $0.1 < \phi < 1.0$.  For every
possible value of $R_s$, we rescale all particle coordinates $\{\mathbf{r}_i\}$
of the specific configuration such that $|\mathbf{r}_i| = R_s$ and ensure that
all rod orientations $\{\hat{\mathbf{u}}_i\}$ lie tangent to the sphere, and
subsequently perform a dynamics simulation at fixed $R=R_s$ for a total
duration of $50\tau$.  We then measure how many particles $N_r$ have undergone
a significant center-of-mass displacement $\Delta r$ during any time interval
$\Delta t$ over the course of this simulation run.  After some testing, we have
found that a suitable stability criterion is $\Delta r = 0.13\lambda$ and
$\Delta t=10\tau$, which corresponds to a displacement of approximately $17\%$
of a rod's width during the time it would take a free rod with $a=2$ to swim
its own length ($\ell = 2\lambda$).  We designate a configuration at a certain
$R_s$ and corresponding packing fraction $\phi$ as stable if and only if
$N_r=0$, and denote the lowest and highest possible packing fractions with
$N_r=0$ as $\phi_{\rm{min}}$ and $\phi_{\rm{max}}$, respectively.

The dependence of the stability on the magnitude of the self-propulsion force,
as shown in Fig.\ \ref{fig:stab3D}, was calculated by first performing a
dynamics simulation of $N=800$ particles with activity strength $F=1$,
undergoing 20 consecutive breathing cycles for $\phi_{\rm{init}}=0.5$ and
$\phi_{\rm{br}}=0.42$.  As above, we placed every particle configuration
formed after a full breathing cycle onto a new sphere with varying radius $R_1
> R_s > R_2$ by rescaling all particle coordinates to $|\bm{r}_i| = R_s$ and
projecting all orientation vectors $\{\hat{\mathbf{u}}_i\}$ to the locally
tangent plane.  For every such set of initial particle coordinates, we equipped
each particle with a constant self-propulsion strength $0 < F < 2.0$ 
and subsequently simulated the dynamics for a time span of $50\tau$.  We
used the same stability criterion as above, $\Delta r = 0.13\lambda$ and
$\Delta t=10\tau$, and deem the system stable if $N_r = 0$.  Note that one
could also introduce a more refined stability criterion that is explicitly
$F$-dependent; however, inspection by eye of the various trajectories for
different $F$-values showed that our current criterion is reasonable for all
cases considered.  Furthermore, it may be seen from Fig.\ \ref{fig:stab3D} that
the resulting stability bounds $\phi_{\rm{min}}$ and $\phi_{\rm{max}}$ vary
non-monotonously with $F$--an important point that would still hold for a
monotonously changing choice of $\Delta r$.


\section{Data availability}
Data are available on request from the authors.

\section{Acknowledgments}
We thank Giorgio Pessot, J\"{u}rgen Horbach, Robert Jack, and David Reichman for helpful discussions. 
L.C.M.J. thanks the Alexander von Humboldt Foundation for support through a Humboldt Research Fellowship. 
A.K. gratefully acknowledges financial support through a Postdoctoral Research Fellowship (KA 4255/1-2) 
from the Deutsche Forschungsgemeinschaft (DFG).
H.L. acknowledges the DFG for support through Science Priority Program SPP1726.

\section{Author contributions}
L.M.C.J.\ and H.L.\ designed research and interpreted the data. 
L.M.C.J.\ developed the simulation code with initial help from A.K. 
L.M.C.J.\ carried out all calculations and wrote the paper,
and all authors commented on the manuscript.

\section{Additional information}
The authors declare no financial interests.

} 



\end{document}